\RequirePackage{fix-cm}
\documentclass[smallextended]{svjour3}       % onecolumn (second format)
\usepackage{graphicx}

\smartqed  % flush right qed marks, e.g. at end of proof

\usepackage{amsfonts, amsmath, amssymb, amsthm, bbm}
\usepackage{enumerate}
\usepackage{natbib}
\usepackage{url} % not crucial - just used below for the URL 
\usepackage{tikz}
\usetikzlibrary{arrows,shapes.arrows,shapes.geometric,shapes.multipart, decorations.pathmorphing,positioning,shapes.swigs}
\usepackage{color}
\definecolor{darkred}{RGB}{100,0,0}
\definecolor{darkgreen}{RGB}{0,100,0}
\definecolor{darkblue}{RGB}{0,0,150}

\usepackage{hyperref}
\hypersetup{colorlinks=true, linkcolor=darkred, citecolor=darkgreen, urlcolor=darkblue}
\usepackage{float}

\newtheorem{thm}{Theorem}
\newtheorem{prp}{Proposition}

\newtheorem{assump}{Assumption}

\theoremstyle{remark}

\newcommand{\eps}{{\varepsilon}}

\newcommand{\E}{\operatorname{\mathbb{E}}}
\renewcommand{\P}{\operatorname{\mathbb{P}}}
\newcommand{\Var}{\operatorname{Var}}

\def\cM{\mathcal{M}}

\journalname{Lifetime Data Analysis}
\begin{document}

%\bibliographystyle{natbib}

%%%%%%%%%%%%%%%%%%%%%%%%%%%%%%%%%%%%%%%%%%%%%%%%%%%%%%%%%%%%%%%%%%%%%%%%%%%%%%
\title{Proximal Causal Inference for Marginal Counterfactual Survival Curves\thanks{Grants or other notes
about the article that should go on the front page should be
placed here. General acknowledgments should be placed at the end of the article.}
}
\subtitle{}

\author{Andrew Ying  \and Yifan Cui \and Eric J. Tchetgen Tchetgen}


\institute{Andrew Ying  \at
              Department of Statistics and Data Science, The Wharton School, University of Pennsylvania \\
           \and
           Yifan Cui \at
           Department of Statistics and Data Science, National University of Singapore
           \and
           Eric J. Tchetgen Tchetgen \at
           Department of Statistics and Data Science, The Wharton School, University of Pennsylvania
                         \email{ett@wharton.upenn.edu}           %  \\
}
\date{Received: date / Accepted: date}

\maketitle

\begin{abstract}
%Treatment-specific survival curves (or equivalently, distribution functions) are important measures with censored time-to-event data that usually arise in biomedical, pharmaceutical and epidemiological studies.
Contrasting marginal counterfactual survival curves across treatment arms is an effective and popular approach for inferring the causal effect of an intervention on a right-censored time-to-event outcome.
%enrich understanding effectiveness and safety of an treatment, which provides more information than commonly investigated central measures such as mean of the survival time.
A key challenge to drawing such inferences in observational settings is the possible existence of unmeasured confounding, which may invalidate most commonly used methods that assume no hidden confounding bias.
%standard approach to estimate causal effects in terms of survival curves in non-experimental settings typically untestable ``no unmeasured confounders (NUC)'' assumption, which is commonly subject to questioning. 
In this paper, rather than making the standard no unmeasured confounding assumption, we extend the recently proposed proximal causal inference framework of \citet{miao2018identifying, tchetgen2020introduction,cui2020semiparametric} to obtain nonparametric identification of a causal survival contrast by leveraging observed covariates as imperfect proxies of unmeasured confounders. Specifically, we develop a proximal inverse probability-weighted (PIPW) estimator, the proximal analog of standard IPW, which allows the observed data distribution for the time-to-event outcome to remain completely unrestricted. PIPW estimation relies on a parametric model for a so-called treatment confounding bridge function relating the treatment process to confounding proxies. As a result PIPW might be sensitive to model mis-specification. To improve robustness and efficiency, we also propose a proximal doubly robust estimator and establish uniform consistency and asymptotic normality of both estimators. We conduct extensive simulations to examine the finite sample performance of our estimators, and proposed methods are applied to a study  evaluating the effectiveness of right heart catheterization in the intensive care unit of critically ill patients.

\end{abstract}

\section{Introduction}
%%% background and motivation
%%% survival curse and its importance in causal inference studies
%Treatment-specific marginal counterfactual survival curves are important measures
%that provide more information than other commonly investigated central measures like mean. 
%Contrasting treatment-specific survival curves can provide insights and understanding in biomedical, pharmaceutical and epidemiological studies that might not be captured by other scales. Moreover, knowledge on survival curves can easily recover other measures like restricted mean survival time or median survival time. Survival curves are especially important in survival studies where right censoring usually accompanies.

%%% NUC and its implausibility
Causal inference about the effects of a potential intervention on a right-censored time-to-event outcome is often conducted by comparing for different treatment values, the marginal counterfactual survival curve for the potential outcome, had possibly contrary to fact, the treatment been set to a given value for the entire population.  
%Kaplan-Meier or Nelson-Aalen survival curves In the case of censored survival data from a randomized trial, a practitioner typically compare treatment-specific Kaplan-Meier or Nelson-Aalen survival curves. In settings where a randomized trial is infeasible due to its cost or unethicality, observational data are frequently used to compare treatment-specific survival. 
However, in observational settings, identification of marginal counterfactual survival curves typically relies on a key assumption of ``no unmeasured confounding (NUC)'', which often presumes that one has accurately measured all relevant common causes of the treatment and outcome variables. In practice, NUC assumption can seldom be ensured to hold, even in well-designed observational studies, and therefore, concern about hidden confounding bias in such studies is warranted.  

%%% IV to handle NUC
To address such confounding concerns, a rich literature has developed over the years on alternative identification strategies for observational settings. Specifically in the survival context, several papers have recently leveraged an instrumental variable (IV) to identify the causal effect of a point treatment on a survival outcome, under either a structural Cox proportional Hazards model or a structural additive hazards model \citep{tchetgen2015instrumental, li2015instrumental,chan2016, martinussen2017instrumental, wang2018learning, ying2019two, cui2021instrumental, ying2021new}. An IV is a pre-treatment variable known to be associated with the treatment variable, to only affect the outcome through its effects on the treatment, and to otherwise be independent of any unmeasured confounder. In addition to these three key IV conditions, a fourth assumption is typically needed for point identification, involving either a functional form restriction on the structural model of interest or a homogeneity condition in the dependence of the treatment variable on the IV and unmeasured confounding factors; neither of which condition can be tested empirically in general.

%%% proximal causal inference framework that handles NUC
More recently, a proximal causal inference (PCI) framework has been proposed to nonparametrically identify causal effects in the presence of unmeasured confounding for which proxies are available under suitable conditions \citep{miao2018identifying,tchetgen2020introduction,cui2020semiparametric,ghassami2021minimax}. The PCI framework has received considerable attention as it has been developed to correct for confounding bias for individuals treatment regimes \citep{qi2021proximal}, categorical confounding \citep{shi2020multiply}, longitudinal studies \citep{deaner2018panel, tchetgen2020introduction, ying2021proximal}, synthetic controls \citep{shi2021theory}, network confounding and homophily bias \citep{egami2021identification}, reinforcement learning \citep{bennett2021proximal} and various generalizations of difference-in-difference estimation \citep{imbens2021controlling}, etc. Nevertheless, to the best of our knowledge, the PCI framework has not been applied to a survival context. %We complement this literature by first considering leveraging PCI into identifying and estimating survival curses, accommodating right censoring for time-to-event studies.

%%% our contributions
The current paper aims to bridge this gap in the literature by developing a PCI approach to identify the causal effect of a point exposure on the marginal counterfactual survival curve of a censored time-to-event outcome by leveraging proxies of unmeasured confounders. For estimation, we first describe a proximal inverse probability-weighted (PIPW) estimator, which identifies the marginal counterfactual survival function for each treatment arm, while allowing the distribution of the time-to-event outcome to remain fully nonparametric. Like standard IPW estimation under unconfoundedness, PIPW relies on estimating a model for the treatment mechanism, which in the proximal setting involves modeling a so-called treatment confounding bridge function posited by the analyst. Consequently, PIPW may be sensitive to estimation bias due to a possibly incorrect model for the treatment confounding bridge function. Thus, to potentially improve robustness and increase efficiency, we also develop a proximal doubly robust (PDR) estimator that incorporates  a model for a so-called outcome confounding bridge function that restricts the conditional distribution for the time-to-event outcome. The PDR estimator is doubly robust in the sense that it remains consistent and asymptotically normal for the marginal counterfactual survival curve, provided that the model for either the treatment or the outcome confounding bridge function is correctly specified, without necessarily knowing which model is correct. %The PIPW and PDR can be readily used in the time-to-event studies handling right censoring by leveraging inverse probability censoring weighting. 
%We prove the uniform consistency and asymptotic normality of PIPW and PDR. This provides us inferential tools on the difference in survival curves both pointwisely and uniformly. 

%%% organization
The remainder of the article is organized as follows. We introduce notation and key assumptions in Section \ref{sec:pre}, where to simplify the exposition, we introduce the PCI framework assuming no censoring of the outcome in view. In Section \ref{sec:est}, we develop our estimators and account for right censoring in Section \ref{sec:censor} via inverse probability-of-censoring weights \citep{robins2000correcting, scharfstein2002estimation}. In Section \ref{sec:asy}, we establish both consistency and asymptotic normality of the proposed estimators, together with corresponding inferential tools. We examine the finite sample performance of our estimators via extensive simulations in Section \ref{sec:simu}. We further apply the proposed estimators to the data application evaluating the joint causal effects of right heart catheterization (RHC) in the intensive care unit of critically ill patients \citep{connors1996effectiveness} in Section \ref{sec:real}. We end the paper with a discussion in Section \ref{sec:dis}. Proofs, additional regularity conditions, and additional results are provided in the appendix.

\section{Preliminaries}\label{sec:pre}
We first prelude our proximal causal inference (PCI) framework. Under PCI, we allow for the presence of unmeasured confounders that determine both the treatment assignment and the outcome variable. We assume that measured covariates, although not sufficient to account for confounding, may nevertheless be leveraged as imperfect proxies of unmeasured confounders, which requires they satisfy certain key conditions we describe below.
%We then ask an investigator to classify the measured covariates into three types of proxies satisfying certain conditions that we later describe. One can then use these proxies to choose to build outcome- and treatment- confounding bridge functions to estimate the marginal survival functions of the time-to-event outcome.

%\subsection{Notation}
Suppose one has observed $n$ i.i.d. copies of data $(T, A, L)$, where $T$ is the time-to-event outcome of interest, %(e.g. time-to-event)
$A$ represents a binary treatment and $L$ are pre-exposure observed covariates. In this Section, we assume $T$ is fully observed to simplify the exposition, however potential for right censoring of $T$ is introduced in Section \ref{sec:censor}. Let $T(a)$, $a \in \{ 1, 0 \}$ denotes the potential outcome \citep{robins1986new, robins1987graphical} that would be observed if the treatment $A$ were, possibly contrary to fact, set to $a$. We make the following standard consistency assumption that $T = T(A)$ almost surely, which links observed outcomes and potential outcomes via the observed treatment indicator. We write $\P$, $\E$, and $\E_n$ as the population probability, the population mean, and the sample average. 

%\begin{rem}
In addition to consistency, under the positivity condition $0 < \P(A = 1|L) < 1$, almost surely,
and the no unmeasured confounding (NUC) assumption %of \cite{robins1986new, robins1987graphical, robins1997causal} is ETT THESE ARE NOT DUE TO ROBINS
that $T(a) \perp A |L$, which requires that $L$ includes all common causes of $T$ and $A$, it is well known that in the absence of censoring, the survival function $\P(T(a) > t)$ is nonparametrically identified from the observed data distribution by the g-formula of \cite{robins1986new}: $\P(T(a) > t) = \E[\P(T > t|a, L)]$. Below, we discuss the proximal causal inference framework which offers an alternative set of identification conditions that allows for nonparametric identification of the counterfactual survival curve $\P(T(a) > t)$, even when NUC is violated.
%\end{rem}

\begin{figure}[H]
\centering
%\documentclass[class=minimal,border=0pt]{standalone}
%\usepackage{tikz}
%\usetikzlibrary{arrows,shapes.arrows,shapes.geometric,shapes.multipart,
%decorations.pathmorphing,positioning,shapes.swigs,}

%\begin{document}

\begin{tikzpicture}
\tikzset{line width=1.5pt, outer sep=0pt,
ell/.style={draw,fill=white, inner sep=2pt,
line width=1.5pt},
swig vsplit={gap=5pt, line color right=red,
inner line width right=0.5pt}};

\node[name=U0, ell, shape=ellipse]{$U$};

\node[name=X0, below=10mm of U0, ell, shape=ellipse]{$X$};

\node[name=Z0, left=15mm of X0, ell, shape=ellipse] {$Z$};

\node[name=W0, right=15mm of X0, ell, shape=ellipse]{$W$};

\node[name=A0, below left=15mm of X0, ell, shape=ellipse]{$A$};

\node[name=Y, below right=15mm of X0, ell, shape=ellipse]{$Y$};

\draw[->,line width=1.0pt,>=stealth](X0) to (Z0);

\draw[->,line width=1.0pt,>=stealth](X0) to (W0);
\draw[->,line width=1.0pt,>=stealth](X0) to (A0);

\draw[->,line width=1.0pt,>=stealth](X0) to (Y);
\draw[->,line width=1.0pt,>=stealth](A0) to (Y);

\draw[->,line width=1.0pt,>=stealth](U0) to (X0);

\draw[->,line width=1.0pt,>=stealth](U0) to (Z0);

\draw[->,line width=1.0pt,>=stealth](U0) to (W0);
\draw[->,line width=1.0pt,>=stealth](U0) to (A0);

\draw[->,line width=1.0pt,>=stealth](Z0) to (A0);
\draw[->,line width=1.0pt,>=stealth](W0) to (Y);

\draw[->,line width=1.0pt,>=stealth](U0) to (Y);

\end{tikzpicture}

%\end{document}
\caption{A directed acyclic graph illustrating treatment- and outcome-confounding proxies.}
\label{fig:dag2}
\end{figure}
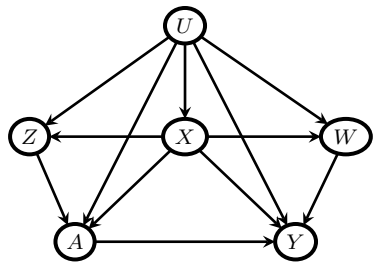

To formally introduce the proximal causal inference framework, suppose now that there are unmeasured variables $U$ that confound the effect of $A$ on $Y$, furthermore, analogous to \cite{cui2020semiparametric, tchetgen2020introduction}, suppose that one can partition the observed covariates $L$ into variables $(X, Z, W)$, where $X$ includes observed common causes of treatment $A$ and outcome variable $T$; $Z$ is associated with $T$ conditional on $A$ only to the extent that it is associated with $U$; and $W$ is associated with A and Z only to the extent that it is associated with $U$.  As $Z$ may either be a cause or an effect of $A$, it will be referred to as treatment confounding proxy; and $W$, which may cause the outcome $T$ will be referred to as outcome confounding proxies. Figure \ref{fig:dag2} provides a Directed Acyclic Graph (DAG) illustrating (X,W,Z) in relation to (A,Y,U) under these conditions. 
%graphically in a setting where NUC holds conditional on $X$, $Z$, and $W$, respectively, so that a g-formula could in principle identify the counterfactual survival curve given the observed data.   

%\begin{figure}[H]
%\centering
%	\begin{minipage}[b]{0.3\linewidth}
%		\centering
%		%\includegraphics[scale=1]{figures/DAG1a.pdf}
%		\resizebox{3.5cm}{!}{\input{figures/DAG1a}}
%		\par Common causes $X$.
%		%\label{fig:dag1a}
%	\end{minipage}
%	\begin{minipage}[b]{0.3\linewidth}
%		\centering
%		%\includegraphics[scale=1]{figures/DAG1b.pdf}
%		\resizebox{3.5cm}{!}{\input{figures/DAG1b}}
%        \par Treatment confounding proxy $Z$.
%		%\label{fig:dag1b}
%	\end{minipage}
%	\begin{minipage}[b]{0.3\linewidth}
%		\centering
%		%\includegraphics[scale=1]{figures/DAG1c.pdf}
%		\resizebox{3.5cm}{!}{\input{figures/DAG1c}}
%		\par Outcome confounding proxy W.
%		%\label{fig:dag1c}
%	\end{minipage}
%\caption{Directed acyclic graphs illustrating treatment- and outcome-inducing proxies.}
%\label{fig:dag1}
%\end{figure}

Formally, $Z$ and $W$ will be said to be valid treatment and outcome confounding proxies if the satisfy the following conditional independence conditions, which we adopt as primitive conditions:
\begin{equation}\label{eq:proxylsi}
    (Z, A) \perp (W, T(a))~\big|~X, U.
\end{equation}
%Figure \ref{fig:dag2} overlays treatment, outcome with treatment and outcome confounding proxies $(Z,W)$ together with observed common causes $X$ and unmeasured confounders $U$, so that the g-formula based on observed data fails to identify the counterfactual survival curve.

Our framework can accommodate any contrast of marginal counterfactual survival curves. In this paper, as an illustration, we focus on the difference in the marginal counterfactual survival curves between the treatment group and the control group, that is,
\begin{equation}\label{eq:estimands}
    \psi(t) = \P(T(1) > t) - \P(T(0) > t).
\end{equation}
%We do remark that since later our estimator is based on estimating the marginal survival function $\P(T(a) > t)$ for one treatment arm $A = a$ first and then take the difference, other measures like marginal survival ratio between groups are also readily estimable by our framework.

%\subsection{Assumptions}
In addition to the consistency assumption: $T = T(A)$ almost surely and positivity: $0 < \P(A = 1|U, X) < 1$ almost surely. We require the following proximal relevance assumptions to ensure existence and uniqueness of the confounding bridge functions introduced in Proposition \ref{prp:iden} below. The assumptions require that $(W, Z)$ are informative about the unmeasured confounders $U$ %ETT THE FOLLOWING IS IMPLIED BY THE FORMER SO DELETE and also each other. 
which formally entails the following so-called completeness conditions :
% ETT THE PAPER NEEDS TO BE SELF CONTAINED, ADD A DISCUSSION OF COMPLETENESS.  DSELETE THE FOLLOWING assumptions can be seen as a variant of completeness assumptions that statisticians are more familiar with. A thorough discussion on examples, histories, and intuition of completeness assumption can be found in \citet{ying2021proximal}.

%\subsection{Proximal Identification}\label{sec:iden}
\begin{assump}[Proximal Relevance for Treatment Confounding Bridge Function]\label{assump:trtcomplete}
For any $a$, $x$ and any square-integrable function $g$,
\begin{enumerate}
    \item $\E[g(U)|W, A = a, X = x] = 0$ almost surely if and only if $g(U) = 0$ almost surely;
    \item $\E[g(W)|Z, A = a, X = x] = 0$ almost surely if and only if $g(W) = 0$ almost surely.
\end{enumerate}
\end{assump}

\begin{assump}[Proximal Relevance for Outcome Confounding Bridge Function]\label{assump:outcomplete}
For any $a$, $x$ and any square-integrable function $g$,
\begin{enumerate}
    \item $\E[g(U)|Z, A = a, X = x] = 0$ almost surely if and only if $g(U) = 0$ almost surely;
    \item $\E[g(Z)|W, A = a, X = x] = 0$ almost surely if and only if $g(Z) = 0$ almost surely.
\end{enumerate}
\end{assump}

The following proposition essentially follows from analogous results from \citet{miao2018identifying, tchetgen2020introduction, cui2020semiparametric}. The main contribution of the result is a nontrivial extension incorporating the time index $t$ into the outcome confounding bridge function $h$ defined in the proposition, a step not required in these prior works.
\begin{prp}\label{prp:iden}
~~~~~
\begin{enumerate}
\item Under Assumption \ref{assump:trtcomplete}(2) and regularity Conditions \ref{assump:idenregularity}(1, 4, 5) in the Appendix, there exists a treatment confounding bridge function $Q(a) = q(Z, a, X)$, that solves the integral equation
\begin{equation}\label{eq:treatconfbridgeiden}
    \frac{1}{\P(A|W, X)} = \E[Q(A)|A, W, X],
\end{equation}
%then, with further assumptions \ref{assump:outproxies}, \ref{assump:trtproxies}, \ref{assump:lnuc} and \ref{assump:trtcomplete}(1), any functions $q$ satisfying \eqref{eq:treatconfbridgeiden}, we have
furthermore, under \eqref{eq:proxylsi} and Assumption \ref{assump:trtcomplete}(1), any such $q$ satisfying \eqref{eq:treatconfbridgeiden}, we have that 
\begin{equation}\label{eq:causaltreatconfbridgegform}
    \P(T(a) > t) = \E[\mathbbm{1}(A = a)Q(A)\mathbbm{1}(T > t)] = \E[\mathbbm{1}(A = a)q(Z, A, X)\mathbbm{1}(T > t)].
\end{equation}
    \item Under Assumption \ref{assump:outcomplete}(2) and regularity Conditions \ref{assump:idenregularity}(1, 2, 3) given in the Appendix, there exists an outcome confounding bridge function $H(t, a) = h(t, W, a, X)$ that solves the integral equation 
\begin{equation}\label{eq:outconfbridgeiden}
    \P(T > t|A, Z, X) = \E[H(t, A)|A, Z, X],
\end{equation}
%then, with further assumptions \ref{assump:outproxies}, \ref{assump:trtproxies}, \ref{assump:lnuc} and \ref{assump:outcomplete}(1), for $h$ satisfying \eqref{eq:outconfbridgeiden}, we have
furthermore, under \eqref{eq:proxylsi} and Assumption \ref{assump:outcomplete}(1), for any function $h$ satisfying \eqref{eq:outconfbridgeiden}, we have that 
\begin{align}
    \P(T(a) > t) &= \E[\mathbbm{1}(A = a)Q(A)(\mathbbm{1}(T > t) - H(t, A)) + H(t, a)] \\
    &= \E[\mathbbm{1}(A = a)q(Z, A, X)(\mathbbm{1}(T > t) - h(t, W, A, X)) + h(t, W, a, X)].\label{eq:causaloutconfbridgegform}
\end{align}
\end{enumerate}
\end{prp}
Proposition \ref{prp:iden} not only defines treatment- and outcome- confounding bridge functions but also suggests two-stage estimators of $\psi(t)$. The idea which is further elaborated in the next Section, proceeds by first estimating the confounding bridge functions $q$ and $h$, and then using equations \eqref{eq:outconfbridgeiden} and \eqref{eq:causaloutconfbridgegform} of the Proposition to construct corresponding estimators for $\psi(t)$. 
\section{Proximal Estimation and Inference}\label{sec:est}
For estimation and inference,
%Parametric \citep{tchetgen2020introduction, cui2020semiparametric, ying2021proximal} or
nonparametric estimation of $q$ and $h$, such as those of \citep{singh2020kernel, ghassami2021minimax, kallus2021causal} could in principle be conducted, however, such methods may not perform well in moderate to high dimensional settings most routinely encountered in practice. Instead, in this paper, for the purpose of illustration, we proceed with the more practical approach of positing parametric model for $q$ and $h$, mainly $q = q(z, a, x; \alpha)$ and $h = h(t, w, a, x; \beta)$ indexed by finite-dimensional parameters $\alpha$ and $\beta$.

%%% First step: estimation of confounding bridge functions
To estimate the treatment confounding bridge function $q$. we propose to solve the following estimating equation for $\alpha$,
\begin{equation}\label{eq:qest}
    \E_n[q(Z, A, X; \alpha)(N - N_+)] = 0,
\end{equation}
where $N = n(W, A, X)$ is a user-specified function of the same dimension as $\alpha$ and $N_+ = n(W, 1, X) + n(W, 0, X)$.

%%% Second step: estimation of survival curves
The resulting proximal inverse probability-weighted (PIPW) estimator is then given by
\begin{equation}\label{eq:pipw}
    \hat \P_{\text{PIPW}}(T(a) > t) = \E_n[\mathbbm{1}(A = a)q(Z, a, X; \hat \alpha)\mathbbm{1}(T > t)].
\end{equation}
And also, inspired by \citet{cui2020semiparametric}, we also propose the proximal doubly robust (PDR) estimator. To that end, we first estimate the outcome confounding bridge function $h$ by solving the following estimating equation for $\beta$,
\begin{equation}\label{eq:hest}
    \E_n\left[\int_0^\infty (\mathbbm{1}(T > t) - h(t, W, A, X; \beta)) dM(t)\right] = 0,
\end{equation}
where $M(t) = m(t, Z, A, X)$ is a user-specified function of the same dimension as $\beta$. The proximal doubly robust (PDR) estimator is then defined as
\begin{align}
    &\hat \P_{\text{PDR}}(T(a) > t) \\
    &= \E_n[\mathbbm{1}(A = a)q(Z, a, X; \hat \alpha)(\mathbbm{1}(T > t) - h(t, W, A, X; \hat \beta)) + h(t, W, a, X; \hat \beta)]. \label{eq:pdr}
\end{align}
As discussed in Section \ref{sec:asy}, this latter estimator has the desirable property of remaining consistent and asymptotically normal for the marginal counterfactual survival curve if either confounding bridge function is correctly specified without necessarily knowing which is mis-specified. Thus far, we have assumed no censoring of the time-to-event outcome, the next section considers the more realistic setting of a censored failure time outcome.  

\subsection{Censoring}\label{sec:censor}
In this Section, we describe a simple approach to account for right censoring of the failure time; in this vein, let $C$ denote a unit's censoring time, which is assumed to satisfy the following standard conditional independence assumption: 
\begin{assump}[Independent Censoring]\label{assump:censoring}
\begin{equation}
    T \perp C ~|~A, X, W, Z.
\end{equation}
\end{assump}
We implement inverse probability-of-censoring weights (IPCW) \citep{robins2000correcting, scharfstein2002estimation} which can appropriately handle censoring under Assumption \ref{assump:censoring}. The idea is to re-weight each unit's contribution to a risk set by the inverse of their probability of remaining uncensored conditional on $(A, L)$. To that end, we require an estimate of $\P(C > t|A, L)$, which can be obtained under say, under a standard Cox proportional hazards model or an Aalen additive hazards model. Throughout the paper, we assume that the censoring distribution can consistently be estimated via such a parametric or semiparametric approach, with yields at each $t$, a regular and asymptotically linear estimator $\hat \P(C > t|A, X, W, Z)$. Note that censoring does not impact estimation of $q$ based on equation \eqref{eq:qest}, however, the PIPW estimator \eqref{eq:pipw} incorporating IPCW becomes
\begin{equation}
    \hat \P_{\text{PIPW}}(T(a) > t) = \E_n\left[\mathbbm{1}(A = a)q(Z, a, X; \hat \alpha)\frac{\mathbbm{1}(T > t, C > t)}{\hat \P(C > t|A, X, W, Z)}\right].
\end{equation}
Likewise, the IPCW version of the estimating equation for $h$ \eqref{eq:hest} becomes
\begin{equation}
    \E_n\left\{\int_0^\infty \left[\frac{\mathbbm{1}(T > t, C > t)}{\hat \P(C > t|A, X, W, Z)} - h(t, W, A, X; \beta)\right]dM(t, Z, A, X)\right\} = 0,
\end{equation}
and the corresponding IPCW PDR \eqref{eq:pdr} is
\begin{align}
    &\hat \P_{\text{PDR}}(T(a) > t) \\
    &= \E_n\Bigg[\mathbbm{1}(A = a)q(Z, a, X; \hat \alpha)\left(\frac{\mathbbm{1}(T > t, C > t)}{\hat \P(C > t|A, X, W, Z)} - h(t, W, a, X; \hat \beta)\right) \\
    &+ h(t, W, a, X; \hat \beta)\Bigg].
\end{align}
Finally, we can define corresponding estimators of additive causal effect on the marginal survival scale as 
\begin{equation}\label{eq:diffpipw}
    \hat \psi_{\text{PIPW}}(t) = \hat \P_{\text{PIPW}}(T(1) > t) - \hat \P_{\text{PIPW}}(T(0) > t),
\end{equation}
and
\begin{equation}\label{eq:diffpdr}
    \hat \psi_{\text{PDR}}(t) = \hat \P_{\text{PDR}}(T(1) > t) - \hat \P_{\text{PDR}}(T(0) > t).
\end{equation}

\subsection{Theorems and Inferences}\label{sec:asy}

Define
\begin{align*}
    {\cal M}_q&=\{\text{all regular laws for the observed data such that } q(Z, A, X) = q(Z, A, X; \alpha),\\
    &\text{ for some value of }\alpha, ~\text{such that \eqref{eq:treatconfbridgeiden} holds}\},
\end{align*}
\begin{align*}
    {\cal M}_h&=\{\text{all regular laws for the observed data such that } h(t, W, A, X)=h(t, W, A, X; \beta),\\
    &\text{ for some value of } \beta, ~\text{such that \eqref{eq:outconfbridgeiden} holds}\}.
\end{align*}
Proofs for the following results are sketched in the Appendix. 

%% Consistency Theorem 
\begin{thm}\label{thm:consistency}
Under \eqref{eq:proxylsi}, Assumptions \ref{assump:trtcomplete}-\ref{assump:censoring} and standard regularity conditions provided in the Appendix, we have that $\sup_t|\hat \psi_{\text{PIPW}}(t) - \psi(t)| \to 0$ and $\sup_t|\hat \psi_{\text{PDR}}(t) - \psi(t)| \to 0$ in probability as $n \to \infty$ under $\cM_q$ and $\cM_q \cup \cM_h$, respectively.
\end{thm}

%% Asymptotic normality Theorem
\begin{thm}\label{thm:asygaussian}
Under \eqref{eq:proxylsi}, Assumptions \ref{assump:trtcomplete}-\ref{assump:censoring} and some standard regularity conditions, we have $\sqrt{n}(\hat \psi_{\text{PIPW}}(t) - \psi(t))$ and $\sqrt{n}(\hat \psi_{\text{PDR}}(t) - \psi(t))$ tend to zero-mean Gaussian processes weakly with covariance process $\Sigma_{\text{PIPW}}(t, s)$ and $\Sigma_{\text{PDR}}(t, s)$ given in the Appendix when $n \to \infty$ under $\cM_q$ and $\cM_q \cup \cM_h$.
\end{thm}
%% Pointwise Inference (immediate by theorem and normal approximation)
Inferences about $\psi(t)$ at a fixed $t$ can be based on the $\alpha$-level Wald confidence interval $(\hat \psi_{\text{PIPW}}(t) + \Phi^{-1}(\alpha/2))\sqrt{\hat \Sigma_{\text{PIPW}}(t, t)}$, $\hat \psi_{\text{PIPW}}(t) + \Phi^{-1}(1 - \alpha/2) \sqrt{\hat \Sigma_{\text{PIPW}}(t, t)})$ and $(\hat \psi_{\text{PDR}}(t) + \Phi^{-1}(\alpha/2))\sqrt{\hat \Sigma_{\text{PDR}}(t, t)}, \hat \psi_{\text{PDR}}(t) + \Phi^{-1}(1 - \alpha/2) \sqrt{ \hat \Sigma_{\text{PDR}}(t, t)})$. Standard error estimates are given by $\hat \Sigma_{\text{PIPW}}(t, t) = \E_n[\hat \eps_{\text{PIPW}}(t)^2]$ and $\hat \Sigma_{\text{PDR}}(t, t) = \E_n[\hat \eps_{\text{PDR}}(t)^2]$, respectively, by writing the influence functions of $\hat \psi_{\text{PIPW}}(t)$ and $\hat \psi_{\text{PDR}}(t)$\footnote{We have $\hat \psi_{\text{PIPW}}(t) = \E_n[\eps_{\text{PIPW}}(t)] + o(1/\sqrt{n})$, $\hat \psi_{\text{PDR}}(t) = \E_n[\eps_{\text{PDR}}(t)] + o(1/\sqrt{n})$. } as $\eps_{\text{PIPW}}(t)$ and $\eps_{\text{PDR}}(t)$ ($\hat \eps_{\text{PIPW}}(t)$ and $\eps_{\text{PDR}}(t)$ at estimated values). 
%The pointwise normal-based confidence intervals at any time $t$ follow immediately. The influence functions, $\Sigma_{\text{PIPW}}(t)$ and $\Sigma_{\text{PDR}}(t)$ are given in the appendix.

%% Uniform Inference (test by resampling)
The causal null hypothesis $\psi(t) = 0$ for all $t$, can be tested by investigating how extreme the test statistic $\sup_{t}|\hat \psi(t)|$ is relative to its distribution under the null. Theorem \ref{thm:asygaussian} does not directly yield this distribution. However, based on the influence functions of our estimators, the distribution of the test statistic $\sup_{t}|\hat \psi(t)|$ under the null can be simulated using the following approach of \citet{lin1993checking}. To that end, suppose one generates 1000 Monte Carlo realizations of $n$ i.i.d. samples from a standard normal variable, such that for the m-th realization $J^m \sim \text{Normal}(0, 1)$, one can show that given the observed data, 
\begin{equation}
    \hat{G}_m(\cdot) = \E_n[\hat \eps(\cdot)J^m],
\end{equation}
converges in distribution to a zero-mean Gaussian process with covariance process $\Sigma(t, s)$. The sampled set $\{\sup_{t}|\hat{G}_m(t)|: 1 \leq m \leq 1000\}$ mimics 1000 simulated samples from $\sup_{t}|\hat \psi(t)|$.
%ETT IM NOT SURE YOU NEED THE SUMMARY BELOW, I WOULD SUGGEST DELETING
%We summarize our estimation and inference procedure as follows:
%\begin{enumerate}
%    \item Estimate the censoring distribution conditional on $(A, X, W, Z)$, namely, $\P(C > t|A, X, W, Z)$;
%    \item Estimate the confounding bridge functions $h(t, W, A, X)$ and $q(Z, A, X)$;
%    \item Estimate the marginal survival functions $\P(T(a) > t)$ for each treatment arm via PIPW, PDR, and take the difference to estimate $\psi(t)$.
%    \item One can use the residuals (or equivalently, the influence functions) $\eps_i(t)$ to draw inferences about $\psi(t)$.
%\end{enumerate}

\section{Simulation}\label{sec:simu}
In this Section, we report a simulation study we conducted to investigate the finite sample performance of proposed methods. Our simulations are based on the following data generating mechanism. 

\noindent
\textbf{Data-generating mechanism: }We generate data $(T, W, A, Z, U, X)$ as followed:
\begin{equation}
    X \sim \max(\text{Normal}(1.1, 0.75^2), 0)\footnote{We manually change a negative value to zero to guarantee a positive hazard later when we generate $T$ via an additive hazards model.}, 
\end{equation}
\begin{equation}
    U \sim \max(\text{Normal}(1.1, 0.75^2), 0), 
\end{equation}
\begin{equation}
    \P(A = 1|X, U) = \frac{1}{1 + \exp(-(0.3 + 0.4 \cdot X - 0.6 \cdot U)},
\end{equation}
\begin{equation}
    Z \sim \text{Normal}(-0.2 - 0.3 \cdot X + 0.65 \cdot U, 0.5^2),
\end{equation}
\begin{equation}
    W \sim \text{Normal}(-0.6 + 0.4 \cdot X + 0.65 \cdot U, 0.5^2),
\end{equation}
\begin{equation}
    T \sim \text{Exponential}(0.1 + 0.6 \cdot A + 0.25 \cdot X + 0.5 \cdot U).
\end{equation}
\begin{equation}
    C \sim \min(\text{Exponential}(0.2), 2).
\end{equation}
%We set a maximum event time $t_{\text{max}} = 2$ where all events above $t_{\text{max}}$ are censored at $t_{\text{max}}$.
The censoring rate generated to be completely independent is 22\%.

\noindent
\textbf{Estimands: }We consider the difference in the marginal counterfactual survival curves between the treatment groups. Specifically, we aim to estimate% (results for $\beta_0$ is in the appendix),
\begin{equation}
    \psi(t) = \P(T(1) > t) - \P(T(0) > t),
\end{equation}
for $t = 0.25, 0.50, 0.75$.

\noindent
\textbf{Methods: }
In the simulation, because censoring is completely independent, we estimate $\P(C > t|A, X, W, Z) = \P(C > t)$ with the standard Kaplan-Meier estimator. In the Appendix, we establish that the specified data generating mechanism implies the following functional forms for  $q$ and $h$
\begin{equation}\label{eq:simuq}
    q(Z, A, X; \alpha) = 1 + \exp\left[(-1)^{1 - A}( \alpha_0 + \alpha_aA + \alpha_z Z + \alpha_xX)\right],
\end{equation}
\begin{equation}\label{eq:simuh}
    h(t, W, A, X; \beta) = \exp(-\beta_0t - \beta_1t^2- \beta_aAt - \beta_wWt - \beta_xX),
\end{equation}
%as working models, where we show that the confounding bridge functions $q$ and $h$ are compatible with the data generating process in the appendix.
For estimation, we set $n(W, A, X) = (-1)^{1 - A}(1, W, A, X)$  in \eqref{eq:qest}, and $m(t, Z, A, X) = (t, Zt, At, Xt, t^2/2)$ in \eqref{eq:hest}.
%\begin{rem}
%The parameterization of $h$ may be overly smooth which overlooks the heterogeneity across time $t$, compared to the semiparametric models typically used under NUC, like the Cox model or the Aalen model. However, this is a simple example just for illustration and one can propose other ways of parameterization, which can readily fit into our framework with minor adjustments.
%\end{rem}
We evaluate the performance of the proposed estimators in situations where either or both confounding bridge functions are mis-specified by considering models based on a nonlinear transformation of observed variables. In particular, each simulated dataset is analyzed using
\begin{itemize}
    \item The PIPW estimator $\widehat \psi_{\text{PIPW}}(t)$ with correctly specified $q$;
    \item The PDR estimator $\widehat \psi_{\text{PDR}}(t)$ with both $q$ and $h$ correctly specified;
    \item The PDR estimator $\widehat \psi_{\text{PDR, WOR}}(t)$ with correctly specified $q$ and incorrectly specified $h$ by using $W^* = \sqrt{|W|} + 1$ instead of $W$ in \eqref{eq:simuh};
    \item The PIPW estimator $\widehat \psi_{\text{PIPW, WIPW}}(t)$ with incorrectly specified $q$, $Z^* = \sqrt{|Z|}$ instead of $Z$ is used in \eqref{eq:simuq};
    \item The PDR estimator $\widehat \psi_{\text{PDR, WIPW}}(t)$ with incorrectly specified $q$ and correctly specified $h$, by using $Z^* = \sqrt{|Z|}$ instead of $Z$ in \eqref{eq:simuq};
    \item The PDR estimator $\widehat \psi_{\text{PDR, BW}}(t)$ with both $q$ and $h$ incorrectly specified. by using $Z^* = \sqrt{|Z|}$ instead of $Z$ in \eqref{eq:simuq}, and $W^* = \sqrt{|W|}$ instead of $W$ in \eqref{eq:simuh};
    \item The standard DR estimator $\widehat \psi_{\text{DR}}(t)$, which assumes NUC based on $L=(X,Z,W)$. The estimator is based on a semiparametric additive hazards regression model for the outcome regression and a logistic regression model for the propensity score.
\end{itemize}
As theorems establish, $\widehat \psi_{\text{PIPW}}(t)$ and $\widehat \psi_{\text{PDR}}(t)$ are expected to perform well in sufficiently large samples. $\widehat \psi_{\text{PDR, WOR}}(t)$ and $\widehat \psi_{\text{PDR, WIPW}}(t)$ are also expected to work well by double robustness. On the other hand, $\widehat \psi_{\text{PIPW, WIPW}}(t)$, $\widehat \psi_{\text{PDR, BW}}(t)$ and $\widehat \psi_{\text{DR}}(t)$ are expected to be biased as they are based on misspecified confounding bridge models.

\noindent
\textbf{Performance measures: }We examine finite-sample performance of estimators by reporting biases, empirical standard errors (SEE), average estimated standard errors (SD), and coverage probabilities of 95\% confidence intervals using $B = 1000$ simulated data sets of size $N = 1000, 2000$. The standard errors of $\widehat \psi_{\text{PIPW}}(t)$, $\widehat \psi_{\text{PDR}}(t)$, $\widehat \psi_{\text{PDR, WOR}}(t)$, $\widehat \psi_{\text{PIPW, WIPW}}(t)$, $\widehat \psi_{\text{PDR, WIPW}}(t)$, $\widehat \psi_{\text{PDR, BW}}(t)$ are estimated via influence functions while that of $\widehat \psi_{\text{DR}}(t)$ is estimated with the bootstrap with number of bootstrap samples equal to $\text{rep} = 200$.
%% simulation results and conclusions
As simulations demonstrate, $\hat \psi_{PIPW}$ and $\hat \psi_{PDR}$ perform well with small biases for all sample sizes considered when all models are correctly specified, confirming our theoretical results. Confidence intervals attain their nominal level with increasing sample size. $\hat \psi_{PIPW, WIPW}$ incurs substantial bias and its standard error estimator is somewhat conservative when the treatment confounding bridge functions is mis-specified, possibly due to instability when fitting $q()$. $\hat \psi_{PDR}$, $\hat \psi_{PDR, WOR}$ and $\hat \psi_{PDR, WIPW}$ remain consistent, confirming the double robustness property. The corresponding variance estimates approach the empirical variance as sample size increases. Both $\hat \psi_{PDR, BW}$ and $\hat \psi_{DR}$ deviate from the truth with large bias when SRA and the underlying union model $\cM_q \bigcup \cM_h$ are incorrect, regardless of sample size, at all time points considered.

\section{Causal Effects of Right Heart Catheterization}\label{sec:real}
%% more info on the dataset
The Study to Understand Prognoses and Preferences for Outcomes and Risks of Treatments (SUPPORT) evaluated the effectiveness of right heart catheterization (RHC) in the intensive care unit of critically ill patients \citep{connors1996effectiveness}. Patients were followed up until 30 days administrative censoring. These data have been re-analyzed in a number of papers in causal inference literature, under a key exchangeability condition on basis of measured covariates \citep{tan2006distributional, vermeulen2015bias, tan2020model, tan2020regularized, cui2019selective}. It also has been analyzed using PCI by \citet{tchetgen2020introduction, cui2020semiparametric, ghassami2021minimax} who focused on estimating the average treatment effect on 30 days survival.

\begin{figure}[H]
    \centering
    \includegraphics[scale = 0.35]{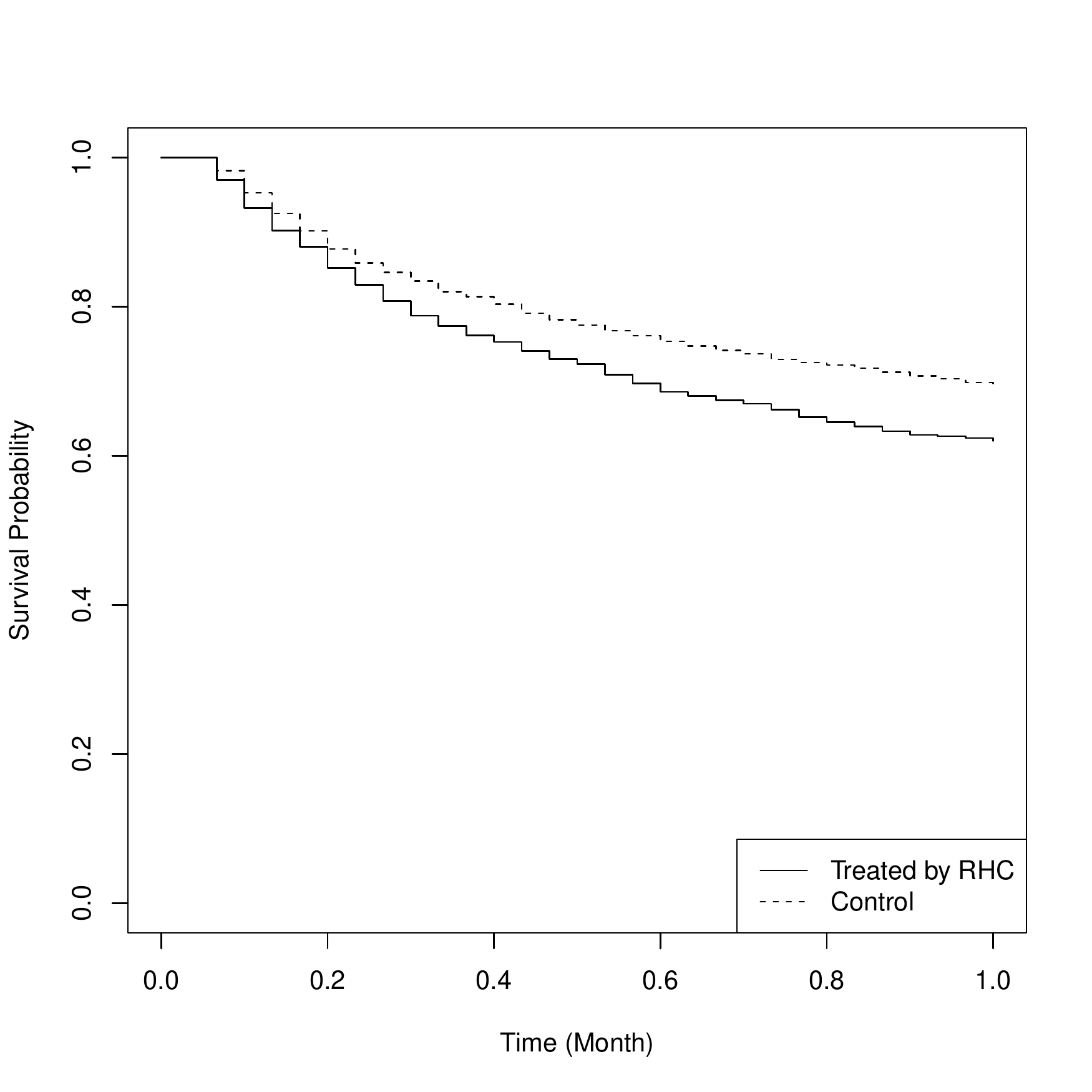}
    \caption{Kaplan-Meier curves of the time to death between groups, where loss to follow up or end of study are treated as censoring.}
    \label{fig:km}
\end{figure}

%% intro to covariates, definition on treatment and outcomes
Data are available on 5735 individuals, with 2184 treated and 3551 controls. In total, 3817 patients survived (1354 treated, 2463 control) and 1918 died (830 treated, 1088 control) within 30 days. The outcome $T$ is the number of days between admission and death or censoring at day 30. We use the month time in the study. Kaplan-Meier curves between two groups are given in Figure \ref{fig:km}. 

As in \citet{tchetgen2020introduction, cui2020semiparametric}, we allocate Z = (pafi1, paco21) and W = (ph1, hema1) and select for X = (age, sex, cat1\_coma, cat2\_coma, dnr1, surv2md1, aps1) as in \citet{ghassami2021minimax}. We specify the outcome confounding bridge function according to \eqref{eq:simuh}, and specify a model of form given by \eqref{eq:simuq} for the treatment confounding bridge function, as in the simulation study. We also fit the data with a standard DR estimator like in the simulation study.

\begin{figure}[H]
    \centering
    \includegraphics[scale = 0.3]{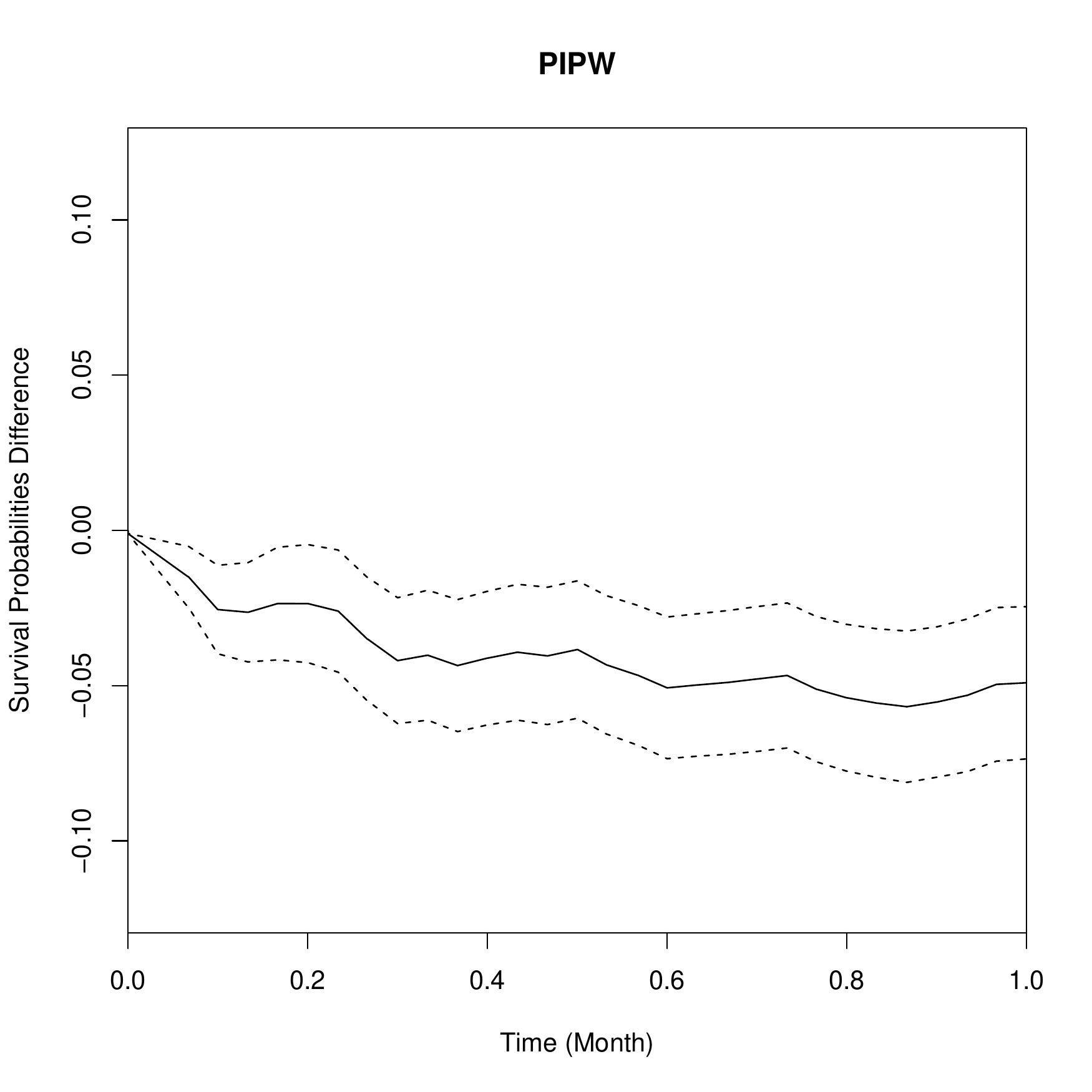}
    \includegraphics[scale = 0.3]{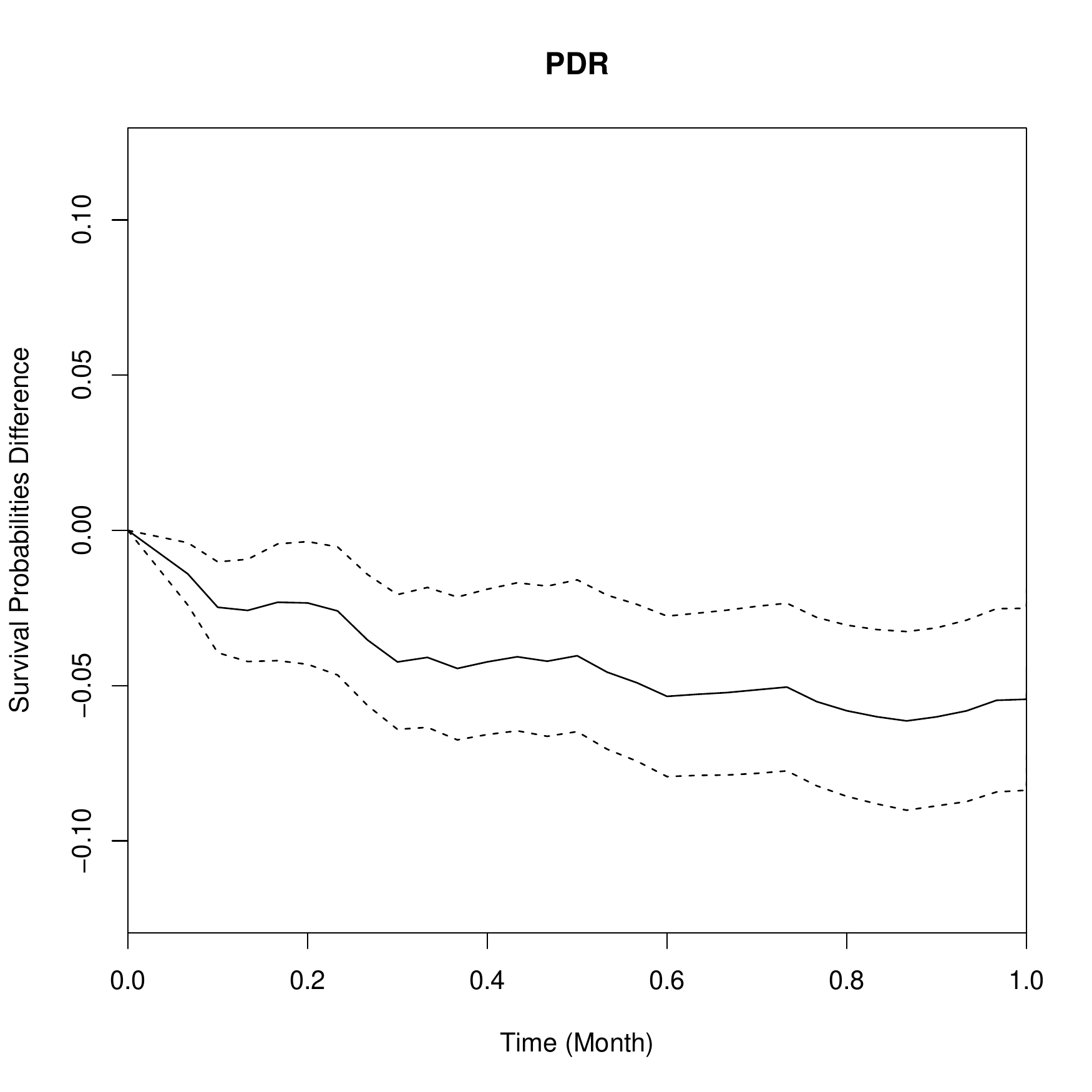}
    \includegraphics[scale = 0.3]{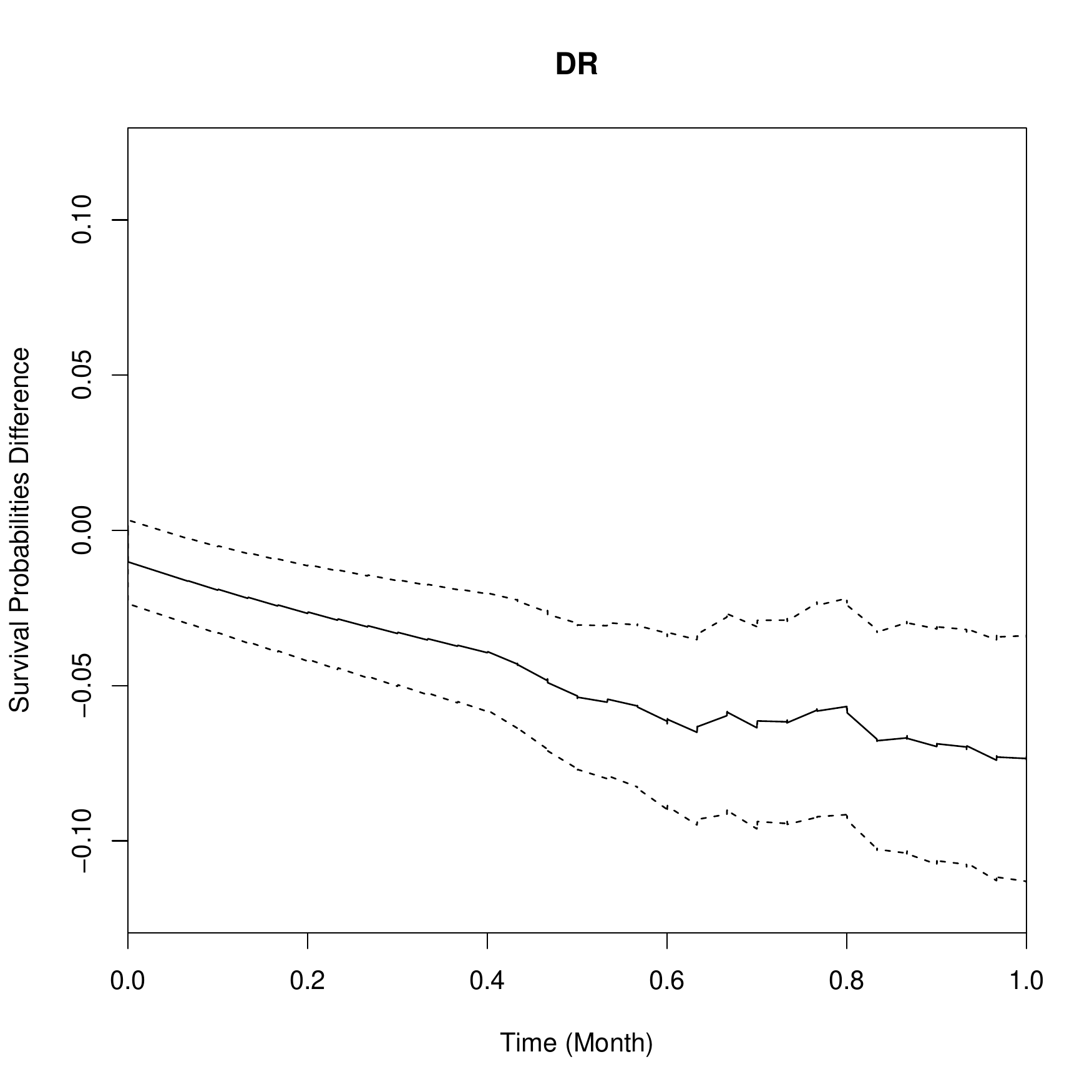}
    \caption{Results of estimated differences in marginal survival functions between treatment and control groups by PIPW, PDR and DR.}
    \label{fig:realresults}
\end{figure}

%% real data analysis results
Estimates of PIPW \eqref{eq:diffpipw}, PDR \eqref{eq:diffpdr} and standard doubly robust (DR) estimator are plotted in Figure \ref{fig:realresults}, together with pointwise normal-based 95\% confidence intervals. The standard deviations used to obtain confidence intervals for PIPW and PDR are estimated via influence functions given in Theorem \ref{thm:asygaussian} while that of $\widehat \psi_{\text{DR}}(t)$ is estimated by the nonparametric bootstrap consisting of  $\text{rep} = 1000$ boostrap samples, as in the simulation study. The significance levels of the supremum tests of the causal null $\psi(t) \equiv 0$ from each estimator are: PIPW: $<$0.001, PDR: $<$0.001, DR: 0.199. PIPW and PDR are close to each other, which suggests that the treatment confounding function is unlikely to be mis-specified. The causal conclusions suggested by supremum tests align with that obtained in \citet{tchetgen2020introduction, cui2020semiparametric, ghassami2021minimax}. These indicate that RHC may have a more harmful effect on survival among critically ill patients admitted into an intensive care unit than previously reported.

\section{Discussion}\label{sec:dis}
%%% a conclusion of our contributions
We have provided a nonparametric identification framework and proposed two estimators for treatment effects on the survival curves scale that do not rely on the ``no unmeasured confounders'' assumption. We have established their asymptotic behavior including uniform consistency and asymptotic normality via standard empirical process theory. The associated inferential tools are developed in R. Empirical evidence supports the asymptotic theory for the proposed estimators and the double robustness property of the proximal doubly robust estimator.

%%% future directions
Our paper provides a first step towards investigating PCI in a survival context. There are many possible future directions. First in this work, we leverage IPCW to handle right censoring, which is known to be inefficient. To improve efficiency, one can in principle augment IPCW to attain local semiparametric efficiency and additional double robustness by providing partial protection against modeling of censoring distribution \citep{rotnitzky2005inverse}. Secondly, although not considered here, the current framework can be generalized to the time-varying treatment and confounding setting of \citep{ying2021proximal}. Another direction that one can also consider is to develop a more robust approach by estimating bridge functions nonparametrically as in \citep{singh2020kernel, ghassami2021minimax, kallus2021causal}. Last, a future direction is to allow more flexible functional form of the function $h(\cdot)$ used in the proximal doubly robust estimator, instead of the parametric form used. For instance, one can estimate $h(\cdot)$ at each event time $t$ to allow coefficients to vary with time $t$.

Finally, a separate interest is to borrow the idea of proximal causal inference to handle informative censoring in survival analysis. By treating right informative censoring as monotone missing not at random and missing indicator as one treatment arm, the proximal causal inference framework with longitudinal data \citep{ying2021proximal} can be directly applied to handle informative censoring, at least, discrete censoring, with minor adjustments. In fact, the independent censoring Assumption \ref{assump:censoring} can be relaxed to informative censoring, as long as the observed covariates $L$ can be decomposed into $(X_C, Z_C, W_C)$ such that $(Z_C, C) \perp (W_C, T) ~|~A, X_C, U_C$, where $X_C$ are common causes of $(T, C)$, $Z_C$ are proxies for censoring process, $W_C$ are proxies for outcome of interest and $U_C$ are unobserved common causes that render censoring time $C$ informative. We illustrate this idea more carefully under discrete censoring with time-varying covariates in the appendix. The extension to continuous censoring is currently being explored by us.

\bibliographystyle{agsm}

\bibliography{ref}

% \usepackage{multirow}
\begin{table}[H]
\caption{\label{tab:simubeta1}
Simulation results of PIPW and PDR estimators. We report bias ($\times 10^{-3}$), empirical standard error (SEE) ($\times 10^{-3}$), average estimated standard error (SD) ($\times 10^{-3}$), and coverage probability of 95\% confidence intervals (95\% CP) of PIPW with correctly specified treatment confounding bridge functions ($\widehat \psi_{\text{PIPW}}(t)$), PDR with both outcome and treatment confounding bridge functions correctly specified ($\widehat \psi_{\text{PDR}}(t)$), PDR with incorrectly specified outcome confounding bridge functions ($\widehat \psi_{\text{PDR, WOR}}(t)$), PIPW with incorrectly specified treatment confounding bridge functions ($\widehat \psi_{\text{PIPW, WIPW}}(t)$), PDR with incorrectly specified treatment confounding bridge functions ($\widehat \psi_{\text{PDR, WIPW}}(t)$), PDR with both outcome and treatment confounding bridge functions wrongly put ($\widehat \psi_{\text{PDR, BW}}(t)$) and a standard doubly robust estimator ($\widehat \psi_{\text{DR}}(t)$) for $\psi_1$, at time $t = 0.25, 0.50, 0.75$, for sample sizes $N = 1000, 2000$ and $B = 1000$ Monte Carlo samples.
}

\centering
%\fbox{
\begin{tabular}{|l|l|ccc|ccc|}
\hline
                                  &$n$ =   & 1000 &     &     & 2000 &       &   \\ \hline
                                &$t$ =  & 0.25 & 0.50 & 0.75     & 0.25 & 0.50 & 0.75   \\ \hline
$\widehat \psi_{\text{PIPW}}(t)$              
&Bias   &-0.4   &0.3    &2.0    &-0.3   &-0.01  &-0.1            \\
&SEE    &32.7   &36.9   &37.1   &22.3   &25.7   &25.7            \\
&SD     &24.6   &23.1   &20.3   &22.8   &26.1   &26.4            \\
&CP     &94.2   &94.7   &94.6   &95.9   &95.5   &95.8            \\\hline
$\widehat \psi_{\text{PDR}}(t)$               
&Bias   &-0.4   &0.3    &2.0    &-0.3   &-0.04  &-0.2             \\
&SEE    &32.6   &36.9   &37.3   &22.4   &25.8   &26.0            \\
&SD     &32.2   &37.1   &37.7   &22.8   &26.2   &26.6            \\
&CP     &94.3   &94.7   &95.3   &95.7   &95.2   &95.9            \\\hline
$\widehat \psi_{\text{PDR, WOR}}(t)$          
&Bias   &-0.4   &0.4    &2.2    &-0.2   &0.1   &0.1            \\
&SEE    &33.1   &37.6   &38.2   &22.3   &26.1   &26.4            \\
&SD     &32.6   &37.7   &38.5   &23.0   &26.6   &27.1            \\
&CP     &94.0   &95.0   &94.7   &95.8   &95.4   &95.3            \\\hline
$\widehat \psi_{\text{PIPW, WIPW}}(t)$        
&Bias   &11.2   &14.7   &18.3   &10.1   &13.7   &14.3            \\
&SEE    &46.4   &49.4   &48.8   &33.0   &36.7   &37.2            \\
&SD     &173    &139    &177    &49.8   &55.8   &58.5            \\
&CP     &93.8   &95.5   &94.5   &92.4   &93.6   &92.3            \\\hline
$\widehat \psi_{\text{PDR, WIPW}}(t)$        
&Bias   &-1.5   &-0.7   &1.2   &-0.5   &-0.1   &-0.8            \\
&SEE    &45.5   &49.3   &49.0   &31.9   &36.1   &36.9            \\
&SD     &172    &141    &179    &50.2   &57.3   &60.0            \\
&CP     &95.9   &97.2   &96.3   &96.8   &95.9   &96.0            \\\hline
$\widehat \psi_{\text{PDR, BW}}(t)$           
&Bias   &1.5    &4.6    &8.2    &1.8    &4.2    &4.6           \\
&SEE    &45.5   &49.3   &49.2   &31.6   &36.1   &37.2            \\
&SD     &173    &146    &188    &47.8   &55.3   &59.8            \\
&CP     &95.3   &97.2   &96.6   &96.7   &95.8   &94.9            \\\hline
$\widehat \psi_{\text{DR}}(t)$               
&Bias   &9.7    &11.5   &13.9   &9.8    &15.1   &18.0            \\
&SEE    &34.7   &37.8   &37.4   &22.0   &25.3   &24.9           \\
&SD     &30.6   &35.0   &35.3   &21.6   &24.7   &24.7            \\
&CP     &89.5   &93.0   &92.0   &92.6   &91.2   &88.8            \\\hline
\end{tabular}%}
\end{table}

\newpage
\appendix
\section{Additional Assumptions}
To provide sufficient conditions for the existence of confounding bridge function, consider the singular value decomposition \citep[Theorem 2.41]{carrasco2007linear} of compact operators to characterize conditions for existence of a solution to Equations \eqref{eq:outconfbridgeiden} and \eqref{eq:treatconfbridgeiden}. Similar conditions were considered by \cite{miao2018identifying} and \cite{cui2020semiparametric} in the point treatment setting. 

Let $L_2\{F(t)\}$ denote the space of all square integrable functions of $t$ with respect to a cumulative distribution function $F(t)$, which is a Hilbert space with inner product $<g_1, g_2> = \int g_1(t)g_2(t)dF(t)$. Define $T_{a, x}$ as the conditional expectation operator: $L_2\{F(w|a, x)\} \to L_2\{F(z|a, x)\}$, $T_{a, x}h = \E[h(W)|z, a, x]$ and let $(\lambda_{a, x, l}, \varphi_{a, x, l}, \phi_{a, x, l})$ denote a singular value decomposition of $T_{a, x}$. That is, $T_{a, x}\varphi_{a, x, l} = \lambda_{a, x, l}\phi_{a, x, l}$. Also Let $T_{a, x}'$ denote the conditional expectation operator: $L_2\{F(z|a, x)\} \to L_2\{F(w|a, x)\}$, $T_{a, x}q = \E[q(Z)|w, a, x]$ and let $(\lambda_{a, x, l}', \varphi_{a, x, l}', \phi_{a, x, l}')$ denote a singular value decomposition of $T_{a, x}'$.
\begin{assump}\label{assump:idenregularity}
~~~
\begin{enumerate}
    \item ~~~ 
\begin{equation}
    \int\int f(w|z, a, x)f(z|w, a, x)dwdz < \infty.
\end{equation}
    \item ~~~
\begin{equation}
    \sup_t\int \P^2(T(a) > t|z, a, x)f(z| a, x)dz < \infty.
\end{equation}
    \item ~~~
\begin{equation}
    \sup_t\sum_{l = 1}^\infty\lambda_{a, x, l}^{'-2}<\P(T(a) > t|z, a, x), \phi_{a, x, l}'>^2 < \infty.
\end{equation}
    \item ~~~
\begin{equation}
    \int \left(\frac{\E[Q(a)|w, a, x]}{f(a|w, x)}\right)^{-2}f(w|a, x)dw < \infty.  
\end{equation}
    \item ~~~
\begin{equation}
    \sum_{l = 1}^\infty\lambda_{a, x, l}^{'-2}<\frac{1}{f(a|w, x)}, ~~\phi_{a, x, l}'>^2 < \infty.
\end{equation}
\end{enumerate}

\end{assump}

\section{Proofs}
\subsection{Identification with Censoring}
We prove that the PIPW estimator identifies $\P(T(a) > t)$.
\begin{align}
    &\E\left(\frac{\mathbbm{1}(T > t, C > t, A = a)q(Z, A, X)}{\P(C > t|A, L)}\right)\\
    &= \E\left(\frac{\mathbbm{1}(A = a)q(Z, A, X)}{\P(C > t|A, L)}\P(T > t, C > t|A, L)\right)\\
    &= \E\left(\frac{\mathbbm{1}(A = a)q(Z, A, X)}{\P(C > t|A, L)}\P(T > t|A, L)\P(C > t|A, L)\right)\\
    &= \E\left(\mathbbm{1}(A = a)q(Z, A, X)\P(T > t|A, L)\right)\\
    &= \E\left(\mathbbm{1}(T > t, A = a)q(Z, A, X)\right)\\
    &= \P(T(a) > t).
\end{align}

The collection of functions, i.e., the class of difference between any two monotone functions, is Donsker. Therefore, Theorem \ref{thm:consistency} and \ref{thm:asygaussian} can be easily proved by standard empirical process theory, Z-estimation theory and some Taylor's expansion \citep{van1996weak}.

Here we give the influence functions of $\widehat \psi_{\text{PIPW}}(t)$, $\widehat \psi_{\text{PDR}}(t)$ at any time $t$. Define $\psi_*(t)$ as the truth. As in the main text, suppose that $q(\cdot)$ and $h(\cdot)$ are parameterized by finite-dimensional parameters $\alpha$ and $\beta$. Also, suppose the censoring model $\P(C > t|A, L)$ is parameterized by $\theta(t)$, which probably involves some NPMLE. Assume their corresponding estimators have representations 
\begin{equation}
    \hat \alpha - \alpha_* = \frac{1}{n}\sum_{i = 1}^n \eps_{\hat \alpha, i} + o_p(1/\sqrt{n}),
\end{equation}
\begin{equation}
    \hat \beta - \beta_* = \frac{1}{n}\sum_{i = 1}^n \eps_{\hat \beta, i} + o_p(1/\sqrt{n}),
\end{equation}
and
\begin{equation}
    \hat \theta(t) - \theta_*(t) = \frac{1}{n}\sum_{i = 1}^n \eps_{\hat \theta, i}(t) + o_p(1/\sqrt{n}).
\end{equation}
Let 
\begin{equation}
    D_{\text{PIPW}}(t) = \frac{\mathbbm{1}(T > t, C > t, A = 1)q(Z, 1, X; \alpha)}{\P(C > t|A = 1, L;\theta)} - \frac{\mathbbm{1}(T > t, C > t, A = 0)q(Z, 0, X; \alpha)}{\P(C > t|A = 0, L;\theta)}
\end{equation}
be the population estimating equation for PIPW, which is a function of $(t, \psi(t), \alpha, \theta)$. Let
\begin{align}
    D_{\text{PDR}}(t) &= \mathbbm{1}(A = a)q(Z, 1, X; \alpha)\left(\frac{\mathbbm{1}(T > t, C > t)}{\hat \P(C > t|A = 1, L; \theta)} - h(t, W, 1, X; \beta)\right) \\
    &~~~- \mathbbm{1}(A = 0)q(Z, 0, X; \alpha)\left(\frac{\mathbbm{1}(T > t, C > t)}{\hat \P(C > t|A = 0, L; \theta)} - h(t, W, 0, X; \beta)\right) \\
    &+ h(t, W, 1, X; \beta) - h(t, W, 0, X; \beta)
\end{align}
be the population estimating equation for PDR, which is a function of $(t, \psi(t), \alpha, \beta, \theta)$. Then the influence functions of PIPW and PDR are
\begin{align}
    &\hat \psi_{\text{PIPW}}(t) - \psi_*(t) = \frac{1}{n}\sum_{i = 1}^n\eps_{\text{PIPW}, i}(t) + o_p(1/\sqrt{n}) \\
    &= \frac{1}{n}\sum_{i = 1}^n\left\{-\E\left(\frac{\partial D_{\text{PIPW}}(t)}{\partial \psi(t)}\right)\right\}^{-1}\left\{D_{\text{PIPW}, i}(t) + \E\left(\frac{\partial D_{\text{PIPW}}(t)}{\partial \alpha}\right)\eps_{\alpha_*, i} + \E\left(\frac{\partial D_{\text{PIPW}}(t)}{\partial \theta(t)}\right)\eps_{\theta_*, i}(t)\right\} + o_p(1/\sqrt{n}),
\end{align}
and
\begin{align}
    &\hat \psi_{\text{PDR}}(t) - \psi_*(t) = \frac{1}{n}\sum_{i = 1}^n\eps_{\text{PDR}, i}(t) + o_p(1/\sqrt{n}) \\
    &= \frac{1}{n}\sum_{i = 1}^n\left\{-\E\left(\frac{\partial D_{\text{PDR}}(t)}{\partial \psi(t)}\right)\right\}^{-1}\left\{D_{\text{PDR}, i}(t) + \E\left(\frac{\partial D_{\text{PDR}}(t)}{\partial \alpha}\right)\eps_{\alpha_*, i}+ \E\left(\frac{\partial D_{\text{PDR}}(t)}{\partial \beta}\right)\eps_{\beta_*, i} + \E\left(\frac{\partial D_{\text{PDR}}(t)}{\partial \theta(t)}\right)\eps_{\theta_*, i}(t)\right\} + o_p(1/\sqrt{n}).
\end{align}

\section{Additional Results for Simulation}
We first prove \eqref{eq:proxylsi}. After intervening $A = a$ in $T$, by the fact that the distribution of $(Z, W)$ does not depend on $A$ and distribution of $(T(a), A, Z, W)$ is determined by $(X, U)$, we have \eqref{eq:proxylsi}. 

Now we prove Assumptions \ref{assump:trtcomplete}, \ref{assump:outcomplete}. It suffices to prove Assumption \ref{assump:trtcomplete}(1), the rest are similar to prove. Suppose there exists a square-integrable function $g$ such that $\E[g(U)|W, A = a, X = x] = 0$ almost surely while, without loss of generality, $g(U) > 0$ on a set $S$ with $\P(S) > 0$. Define the set of $w$ that corresponds to the possible values, which follows from the fact that $U$ and $W$ are absolutely continuous to each other. We have
\begin{equation}
    \E[g(U)|w, A = a, X = x] = \int g(u)f(u|w, a, x)du > 0.
\end{equation}

Finally we prove that the true $q$ and $h$ can be parameterized in the form of \eqref{eq:simuq} and \eqref{eq:simuh}. A similar proof has shown that there exists an $\alpha$ such that
\begin{equation}
    \E[q(Z, A, X; \alpha)|A, U, X] = \frac{1}{\P(A|U, X)}.
\end{equation}
We show that there exists $\beta$ such that
\begin{equation}
    \E[h(t, W, A, X; \beta)|A, U, X] = \P[T > t|A, U, X].
\end{equation}
Note that the RHS is equal to
\begin{equation}
    \exp[-(0.1 + 0.6 \cdot A + 0.25 \cdot X + 0.5 \cdot U)t - 0t^2],
\end{equation}
whereas the LHS equals
\begin{align}
    &\exp[-(\beta_0 + \beta_aA + \beta_xX)t - \beta_1t^2]\E[\exp[-(\beta_wW)t]|A, U, X]\\
    &=\exp[-(\beta_0 + \beta_aA + \beta_xX + \beta_w\E(W|A, U, X))t - (\beta_1 + 0.5\beta_w^2\Var(W|A, U, X))t^2].
\end{align}
Now since $\E(W|A, U, X)$ is linear in $(A, U, X)$ and $\Var(W|A, U, X)$ is constant in $(A, U, X)$, we have five unknown parameters for five equations. Thus we have a unique solution.

\section{Proximal Causal Inference Applied to Handle Discrete Informative Censoring}
The independent censoring Assumption \ref{assump:censoring} can be relaxed even to informative censoring, by borrowing the idea of proximal causal inference, at least with discrete censoring. In fact, as commonly known in the literature, right informative censoring can be treated as monotone missing not at random \citep{van2003unified, rotnitzky2005inverse, tsiatis2006semiparametric}. Therefore, by treating the observed indicator as one ``treatment arm'' in causal inference and evaluating the expected mean of the counterfactual outcome with intervention to that arm, the proximal causal inference framework \citep{ying2021proximal} can be directly extended to handle informative censoring in longitudinal studies.

Suppose that one has observed $n$ i.i.d. copies of survival data $(T = \min(\tilde T, C), \Delta = \mathbbm{1}(\tilde T < C), \overline L(K))$, where $C$ takes value in $1, 2, \cdots, K$ and we assume there is a maximum visiting time $K$ so that $\P(T \leq K) = 1$. We have observed covariates $L(k)$ provided $T \geq k$. We consider informative censoring, that is, there exists unmeasured time-varying covariates $U(k)$ such that,
\begin{equation}
    \mathbbm{1}(\tilde T > k) \perp \mathbbm{1}(C > k)~|~T \geq k, \bar L(k), \bar U(k).
\end{equation}
We are interested in identifying 
\begin{equation}
    \E(\nu(\tilde T)|V),
\end{equation}
for a known function $\nu()$ with some baseline covariates $V \subset X(0)$. Like in PCI, we assume that the observed covariates $\overline L(k)$ consists of three bucket types $(\overline X(k), \overline Z(k), \overline W(k))$, where $\overline X(k)$ are common causes of subsequent censoring and outcome variables, $C(k), C(k+1), \cdots, C(K)$ and $Y$, $\overline Z(k)$ are referred to as censoring-inducing proxies (type 2) and $\overline W(k)$ are referred to as outcome-inducing proxies (type 3). For identification, we assume proximal independence
\begin{equation}
    (\overline Z(k), \mathbbm{1}(\tilde T > k)) \perp (\mathbbm{1}(C > k), \overline W(k)) ~|~ T \geq k, \overline U(k), \overline X(k).
\end{equation}
We write $C(k) = \mathbbm{1}(C \leq k)$. Note that when $T \geq k$, $C(k) = \mathbbm{1}(T = k)(1 - \Delta)$. With completeness assumption, that is, for any $k, \overline X(k)$, and any square-integrable function $\nu$,
\begin{equation}
    \E[\nu(\overline W(k))|\mathbbm{1}(C \leq k), T \geq k, \overline Z(k), \overline X(k)]~\text{if and only if}~\nu(\overline W(k)) = 0~\text{almost surely}.
\end{equation}
and some regularity conditions, there exist functions $Q_k(c(k)) = q_k(\overline Z(k), c(k), \overline X(k))$ such that (with $Q_0() \equiv 0$)
\begin{equation}\label{eq:treatconfbridgeiden2}
    \E[Q_{k}(C(k))|C(k), T \geq k, \overline W(k), \overline X(k)] = \frac{\E[Q_{k - 1}(0)|T \geq k, \overline W(k), \overline X(k)]}{\P(C(k)|T \geq k, \overline W(k), \overline X(k))},
\end{equation}
for any $0 \leq k \leq K - 1$. This can be shown to imply
\begin{align}
    &\E[\E\{Q_{k}(C(k))|C(k), C(k - 1) = 0, \overline U(k), \overline X(k)\}|C(k), T \geq k, \overline W(k), \overline X(k)\}\\
    &=\E[\E\{Q_{k}(C(k))|C(k), T \geq k, \overline W(k), \overline U(k), \overline X(k)\}|C(k), T \geq k, \overline W(k), \overline X(k)\}\\
    &=\E\{Q_{k}(C(k))|C(k), T \geq k, \overline W(k), \overline X(k)\}\\
    &=\frac{\E[Q_{k - 1}(0)|T \geq k, \overline W(k), \overline X(k)]}{\P(C(k)|T \geq k, \overline W(k), \overline X(k))}\\
    &=\int\frac{\E\{Q_{k - 1}(0)|T \geq k, \overline W(k), \overline U(k), \overline X(k)\}f(\overline U(k)|T \geq k, \overline W(k), \overline X(k))}{\P(C(k)|T \geq k, \overline W(k), \overline X(k))}d\overline U(k)\\
    &=\int\frac{\E\{Q_{k - 1}(0)|T \geq k, \overline W(k), \overline U(k), \overline X(k)\}}{\P(C(k)|T \geq k, \overline W(k), \overline U(k), \overline X(k))}f(\overline U(k)|C(k), T \geq k, \overline W(k), \overline X(k))d\overline U(k)\\
    &=\E\left\{\frac{\E\{Q_{k - 1}(0)|T \geq k, \overline W(k), \overline U(k), \overline X(k)\}}{\P(C(k)|T \geq k, \overline W(k), \overline U(k), \overline X(k))}|C(k), T \geq k, \overline W(k), \overline X(k)\right\}\\
    &=\E\left\{\frac{\E\{Q_{k - 1}(0)|C(k - 1) = 0, \overline U(k), \overline X(k)\}}{\P(C(k)|C(k - 1) = 0, \overline U(k), \overline X(k))}|C(k), T \geq k, \overline W(k), \overline X(k)\right\},
\end{align}
which, by another completeness assumption, that is, for any $k, \overline X(k)$, and any square-integrable function $\nu$,
\begin{equation}
    \E(\nu(\overline U(k))|\mathbbm{1}(C \leq k), T \geq k, \overline W(k), \overline X(k))~\text{if and only if}~\nu(\overline U(k)) = 0~\text{almost surely},
\end{equation}
yields
\begin{equation}
    \E\{Q_{k}(C(k))|C(k), C(k - 1) = 0, \overline U(k), \overline X(k)\} = \frac{\E\{Q_{k - 1}(0)|C(k - 1) = 0, \overline U(k), \overline X(k)\}}{\P(C(k)|C(k - 1) = 0, \overline U(k), \overline X(k))}.
\end{equation}
Then these bridge functions $q_k(\cdot)$ can be used to identify
\begin{align}
    &\E(\nu(\tilde T)|V) = \E[\nu(T)\Delta Q_{\max_{k \leq T}k}(0)|V] = \E[\nu(T)\Delta q_{\max_{k \leq T}k}(\overline Z(k), 0, \overline X(k))|V]\\ 
    &= \E[\sum_{k = 0}^{K - 1}\mathbbm{1}(k < T \leq k + 1)\nu(T)\Delta q_k(\overline Z(k), 0, \overline X(k))|V].
\end{align}
Estimation and inferential approaches can be similarly developed as in \citep{ying2021proximal}.
\end{document}